\documentclass{ws-procs9x6}

\newcommand{\OM}{\Omega_M}

\newcommand{\OMo}{\Omega_{M}^0}

\newcommand{\OLo}{\Omega_{\Lambda}^0}

\newcommand{\OXo}{\Omega_{X}^0}

\newcommand{\OD}{\Omega_{D}}
\newcommand{\ODo}{\Omega_{D}^0}

\newcommand{\wX}{\omega_X}

\newcommand{\CC}{\Lambda}

\newcommand{\we}{\omega_e}

\newcommand{\tOM}{\tilde{\Omega}_M}

\newcommand{\tOD}{\tilde{\Omega}_{D}}

\def\de{\delta}

\def\hh{\hat{h}}

\def\csd{c_{s}^2}
\def\cad{c_{a}^2}

\newcommand{\JHEP}[3]{ {JHEP} {#1} (#2)  {#3}}
\newcommand{\NPB}[3]{{\sl Nucl. Phys. } {\bf B#1} (#2)  {#3}}
\newcommand{\NPPS}[3]{{\sl Nucl. Phys. Proc. Supp. } {\bf #1} (#2)  {#3}}
\newcommand{\PRD}[3]{{\sl Phys. Rev. } {\bf D#1} (#2)   {#3}}
\newcommand{\PLB}[3]{{\sl Phys. Lett. } {\bf B#1} (#2)  {#3}}

\newcommand{\PR}[3]{{\sl Phys. Rep. } {\bf #1} (#2)  {#3}}
\newcommand{\RMP}[3]{{\sl Rev. Mod. Phys. } {\bf #1} (#2)  {#3}}

\newcommand{\MPLA}[3]{{\sl Mod. Phys. Lett. } {\bf A#1} (#2) {#3}}

\newcommand{\JCAP}[3]{{ JCAP} {\bf#1} (#2)  {#3}}
\newcommand{\APJ}[3]{{\sl Astrophys. J. } {\bf #1} (#2)  {#3}}

\newcommand{\MNRAS}[3]{{\sl Mon. Not. Roy. Astron. Soc.} {\bf #1} (#2)  {#3}}

\newcommand{\ProgS}[3]{{\sl Prog. Theor. Phys. Supp.} {\bf #1} (#2)  {#3}}
\newcommand{\APJS}[3]{{\sl Astrophys. J. Suppl.} {\bf #1} (#2)  {#3}}

\begin{document}

\title{DARK ENERGY PERTURBATIONS AND A POSSIBLE SOLUTION TO THE COINCIDENCE PROBLEM}

\author{JAVIER GRANDE\footnote{Speaker}, ANA PELINSON\footnote{Present address: IAG, Univ. de S\~ao Paulo, Rua do Mat\~ao, CEP 05508-900,
S.P., Brazil.} and JOAN SOL\`A}

\address{High Energy Physics Group, Dept. ECM, and Institut de Ci{\`e}ncies del Cosmos\\
    Univ. de Barcelona, Av. Diagonal 647, E-08028 Barcelona, Catalonia, Spain\\
    E-mails: jgrande@ecm.ub.es, apelinson@ecm.ub.es, sola@ifae.es}

\begin{abstract}
We analyze some generic properties of the dark energy (DE)
perturbations, in the case of a self-conserved DE fluid. We also
apply a simple test (the ``F-test'') to compare a model to the data
on large scale structure (LSS) under the assumption of negligible DE
perturbations. We exemplify our discussions by means of the
$\CC$XCDM model, showing that it provides a viable solution to the
cosmological coincidence problem.
\end{abstract}


\bodymatter

\section{Introduction}\label{grande:sec1}

In recent times Cosmology has become an accurately testable branch
of Physics. Theoretical models can now be confronted with a large
quantity of high-precision data coming from different sources,
including studies of distant supernovae \cite{Supernovae}, the
anisotropies of the CMB \cite{WMAP3} or the LSS of the Universe
\cite{Cole05}. All these observations give strong support to the
existence of DE, although the ultimate nature of this component
remains a complete mystery.

Remarkably enough, the simplest DE candidate, namely a cosmological
constant (CC) $\CC$, gives rise to a model ($\CC$CDM) in accurate
agreement with all the currently available observational data.
Moreover, a general prediction of quantum field theory (QFT) is the
existence of a vacuum energy which would precisely take the form of
a CC in the Einstein equations. However, the value predicted by the
theory happens to be many orders of magnitude larger than the
observed DE density. This fact, known as the ``cosmological constant
problem'' \cite{CCRev}, makes very unlikely the identification of DE
with a strictly constant vacuum energy density, since that would
require an extremely fine-tuned cancellation of the different
contributions, unless there is a dynamical mechanism taking care of
such adjustment \cite{oldrelax}. In a different vein, the CC
problem can also be addressed in quantum cosmology models of
inflation, through the idea of multiverses\,\cite{Linde87} and
the application of anthropic considerations \cite{CCRev}.

The CC problem could be alleviated if we allow the DE to be
dynamical. The most popular models exploiting this idea are undoubtedly
the scalar field models (XCDM)\cite{quintessence}.
These models, although well-motivated from the Particle Physics
point of view, present two major drawbacks. First of all, the field
should have an extremely tiny mass, $m_X\sim H_0\sim 10^{-33}$ eV,
which is even much smaller than the observed value of the
mass scale associated to the DE ($\sqrt[4]{\rho_{D}}\sim 10^{-3}$
eV). And second, in this kind of models one implicitly assumes that
the vacuum energy predicted by QFT cancels out for some reason, so
the fine-tuning problems associated to the vacuum energy
are not solved but simply obviated and traded for those of the
scalar field itself. In short, the situation is as
follows: on the one hand, from QFT we expect a vacuum energy
contribution to the DE in the form of a CC; on the other, we have
the popular and well-motivated scalar field models, which may serve
to alleviate the CC problem due to their dynamical nature.

Therefore, it seems quite natural to study a more complete
model in which the DE combines both ingredients, which we call the
$\CC$XCDM model \cite{LXCDM12}. The new model presents additional
advantages: e.g. $\CC$ need not be constant, but may evolve with a
renormalization group (RG) equation, as any other parameter in QFT.
The other DE component, the ``cosmon'' $X$, need not be a
fundamental field either; it could be e.g. an effective
representation of dynamical fields of various sorts or even of
higher order curvature terms in the action. In fact we do not have
to assume anything about the nature of $X$: its dynamics is
completely determined from that of $\CC$ once we have a
good ansatz for the latter, due to the fact that both components may
exchange energy. In the original $\CC$XCDM
model\cite{LXCDM12}, the result of these assumptions is a
3-parameter cosmological framework which incorporates the $\CC$CDM
and XCDM models as special cases.

One of the most appealing features of the $\CC$XCDM model is that it
provides a solution to the ``cosmological coincidence
problem''\cite{CCRev}, i.e. the problem of explaining why the energy
densities of matter and DE are currently of the same order. In the
standard $\CC$CDM model, this fact is indeed a coincidence since the
evolution of the two components is very different; namely, while the
DE density remains constant, the matter density decays fast with the
scale factor as $a^{-3}$. In contrast, in the $\CC$XCDM model, the
ratio $r=\rho_D/\rho_M$ between the DE and matter densities may be
bounded and not vary too far away from $1$ for a significant
fraction of the history of the Universe. It means that, for a very
long time, $\rho_D$ and $\rho_M$ stay naturally of the same order as
they are nowadays. Recently, a generalized version of the $\CC$XCDM
model has been suggested in Ref.~\refcite{BSS09} with even more far
reaching consequences, namely it is able to relax the value of the
DE in the present Universe starting from an arbitrary value in the
early epochs, i.e. it constitutes an interesting attempt to solve
the old CC problem without using the traditional adjustment
mechanisms based on scalar fields\,\cite{oldrelax}.

Whatever its nature, if the DE is not a strict CC, then, according
to cosmological perturbation theory, it should fluctuate. In order
to find out its impact on the LSS formation, we will discuss some
generic properties of the DE perturbations, exemplifying them by
means of the $\CC$XCDM model. We will also address the question of
how the LSS data can be used to constrain a model. The information
about LSS is encoded in the galaxy fluctuation power spectrum,
$P_{GG}(k)$, which is determined observationally and must be
reproduced by the predicted matter power spectrum $P(k)\equiv
|\delta_M(k)|^2$ of the theoretical model. A first, and economical,
approach to the problem is to simply neglect DE perturbations. Using
the fact that the $\CC$CDM model provides a good fit to the data, we
may take it as a reference and impose that the power spectrum of our
model does not deviate by more than $10\%$ from the $\CC$CDM value
(``F-test''\cite{Ftest,GOPS}). As we will see, this simple analysis
may serve to strongly restrict the parameter space of a model.
Nevertheless, it does not reflect some important features that only
come to light when making a full study of the combined system of
matter and DE perturbations. We will show that such a
study\,\cite{GPS} is useful not only to check the validity of the
previous approach, but it can also help us to further constrain
the physical region of the parameter space.

The net result of our analysis of the DE perturbations and its
implication on LSS formation is quite rewarding, as we are able to
find a sizable region of the $\CC$XCDM parameter space where the
model is in full agreement with LSS data, and other cosmological
observations, while providing at the same time a plausible dynamical
solution to the cosmological coincidence problem.

\section{Dark energy perturbations}\label{sec:2}

In this section, we discuss some general properties of the DE
perturbations for models in which both matter and DE are
self-conserved:
\begin{equation}\label{conservation}
\rho'_n+\frac{3}{a}(1+\omega_n)\rho_n=0
\end{equation}
Here a prime denotes differentiation with respect to the scale
factor ($f'\equiv df/da$) and $n=M,D$ stands for each of the energy
components, matter/radiation and DE. We take $\omega_M=0$,
since we are interested in studying the perturbations in the
matter-dominated (MD) era, and we denote the equation of state (EOS) of the DE component
as $\omega_e$, where the subindex ``$e$'' serves us to remember that the
EOS may be an effective one. Let us first introduce the basic
equations for the fluctuations, which we derive following the
standard approach \cite{Kodama84}. For the background space-time we
adopt the spatially flat FLRW metric,
$ds^2=dt^2-a^2\delta_{ij}dx^idx^j$. We perturb it
\begin{equation}
g_{\mu\nu}\rightarrow g_{\mu\nu}+h_{\mu\nu}\,,
\end{equation}
keeping only the scalar part of the perturbation. In order to have
uniquely defined fluctuations $h_{\mu\nu}$, a gauge choice is
mandatory, i.e. we have to choose a specific coordinate system. Here
we adopt the synchronous gauge\cite{Kodama84}, for which
$h_{00}=h_{0i}=0$.

We should also perturb the energy-momentum tensor, considering thus
perturbations on the density, pressure and 4-velocity of each fluid:
\begin{equation}
\rho_n \rightarrow \rho_n+\delta\rho_n\,,\qquad p_n \rightarrow
p_n+\delta p_n\,,\qquad U^\mu_n \rightarrow U^\mu_n+\delta
U^\mu_n\,.
\end{equation}
The equations for the fluctuations are then obtained by perturbing
the 00-component of the Einstein equations,
$R_{\mu\nu}-g_{\mu\nu}R/2=8\pi G T_{\mu\nu}$, and the conservation
law for the energy momentum-tensor, $\nabla_{\mu}T^\mu_\nu=0$. At
the end we obtain 5 equations depending on the following set of 7
variables:
\begin{equation}
\hh \equiv  \frac{\partial}{\partial
t}\left(\frac{h_{ii}}{a^2}\right),\;\quad{\theta}_n\equiv
\,\nabla_\mu (\de U_n^\mu)=\nabla_j (\de
U_n^j),\;\quad\delta_n\equiv\frac{\delta\rho_n}{\rho_n},\;\quad
\delta p_n
\end{equation}
Therefore, in order to solve our system we need to give an
expression for $\delta p_D$ (since for the matter component, indeed
$\delta p_M=0$). In the case of adiabatic perturbations, we simply
have $\delta p_D=c^2_a\delta\rho_D$, where
\begin{equation}
c^2_a=\frac{p'_D}{\rho'_D}=\we-\frac{a}{3}\frac{\we'}{(1+\we)}
\label{cadas}
\end{equation}
is the adiabatic speed of sound of the DE fluid. In general,
however, there could be an entropy contribution to the pressure
perturbation.
In this case, the relation between $\delta p_D$ and
$\delta\rho_D$ in an arbitrary system of reference is given as
follows\,\cite{Kodama84}
\begin{equation}\label{pentropy}
{\de p_D}= \csd{\de
\rho_D}-a^3\,\rho'_D\,H(\csd-\cad)\,\frac{\theta_D}{k^2}\,,
\end{equation}
where $\csd$ is the rest-frame (or \emph{effective}) speed of sound
and $k$ is the wave number, as we have moved to Fourier space. This
expression is gauge-invariant, and thus it can be computed in any
desired gauge, in particular in the synchronous one. When
$\rho_D$ is self-conserved, equation (\ref{conservation}) holds for
$\rho_D$ and, in such case, (\ref{pentropy}) takes on the
form
\begin{equation}
{\de p_D}= \csd{\de
\rho_D}+3Ha^2(1+\omega_e)\rho_D(\csd-\cad)\frac{\theta_D}{k^2}\,.
\end{equation}
Finally, the equations for the perturbations read:
\begin{eqnarray}
&&\hh' + \frac{2}{a} \hh -  \,\frac{3H}{a} \tOM\de_M=
\,\frac{3H}{a}\tOD \left[(1+3\csd)\de_D + 9a^2
H(\csd-\cad)\frac{\theta_D}{k^2} \right]\label{ode}\\
&&\de_M^\prime =
-\frac{1}{aH}\left(\theta_M-\frac{\hh}{2}\right)\\
&&\theta_M^\prime = - \frac{2}{a}\theta_M\\
&&\de^{\,\prime}_D = - \frac{(1+\we)}{aH}\left\{\left[1+\frac{9a^2H^2(\csd-\cad)}
{k^2} \right]\theta_D-\frac{\hh}{2}\right\}-\frac{3}{a}(\csd-\we)\de_D\qquad \label{ode1}\\
&&{\theta}^{\prime}_D=-\frac{1}{a}\left(2-3\csd\right){\theta}_D+\frac{k^2}{a^3H
} \frac{\csd\de_D}{(1+\we)}\,, \label{ode2}
\end{eqnarray}
where $\tilde{\Omega}_n(a)\equiv\Omega_n(a)H_0^2/H^2$ and
$\Omega_n(a)\equiv\rho_n(a)/\rho^0_c=8\pi G\rho_n(a)/(3
H_0^2)$. From these equations we get a second-order differential
equation\,\footnote{In (\ref{secm}) we have corrected a typo that
appears in Eq.\,(50) of Ref.\,\refcite{GPS}.} for $\delta_M$:
\begin{eqnarray}
\delta_M''(a)&+&\frac{3}{2}\left[1-\we (a)\tOD(a)
\right]\frac{\delta_M'(a)}{a}- \frac{3}{2}\,\tOM(a)\
\frac{\delta_M(a)}{a^2}=\nonumber\\
&=&\frac{3}{2}\, \tOD(a) \left[(1+3\csd)\frac{\de_D(a)}{a^2} +
9H(a)\,(\csd-\cad)\frac{\theta_D(a)}{k^2} \right]\,.\label{secm}
\end{eqnarray}
In order to study the
properties of DE perturbations, it is useful to write also a
second-order differential equation for $\delta_D$. This equation is
much simpler if we use differentiation with respect to the conformal
time $\eta$ ($dt=a d\eta$) and work in the comoving gauge. Notice
that gauge issues are unimportant for sub-Hubble perturbations, as
the ones we study here, so the behavior of the perturbations will
not depend on the chosen gauge. Defining the expansion rate in the
conformal time $\mathcal{H}=(da/d\eta)/a\equiv\dot{a}/a$, the
counterpart of Eq.\,(\ref{secm}) in the comoving gauge reads
\cite{Kodama84}:
\begin{equation}
\ddot{\Delta}-\Big[3\left(2\omega_e-c_a^2\right)-1\Big]\mathcal{H}
\dot{\Delta}+3\Bigg[\left(\frac{3}{2}\omega_e^2-4\omega_e-\frac{1}{2}+3c_a^2\right)\mathcal{H}^2+\frac{k^2}{3}c_s^2\Bigg]\Delta=0\,.
\label{secd}
\end{equation}

\subsection{Generic properties of the DE perturbations}\label{sec:2b}

Once we have shown the basic equations, let us discuss some general
properties of the DE perturbations. Looking at \eref{secd}, we see
that the coefficient of $\Delta$ presents two terms. If it is the
second of these terms (the one proportional to $k^2$) that
dominates, and forgetting for a moment about the term proportional
to $\dot{\Delta}$, we are left with the equation of a harmonic
oscillator. This defines the \emph{sound horizon}, a ``Jeans scale''
for the DE,
\begin{equation}
\lambda_{s}=\left|\int^{\eta}_0 c_{s}
\mathrm{d}\eta\right|\label{po}\,,
\end{equation}
such that for scales well inside it, i.e. $l\sim k^{-1}\ll
\lambda_s$:
\begin{equation}
\de_D=C_1 e^{i c_{s}k\eta}+C_2 e^{-i c_{s} k\eta}\,,
\end{equation}
where $C_1$ and $C_2$ are constants, and we have assumed constant $\csd$ for simplicity. Therefore, we
see that:

\begin{itemlist}
\item If $c_s^2<0$, the perturbations grow exponentially, situation
which is unacceptable for structure formation. As long as $\we$ is not
varying too fast, $c^2_a<0$ [cf. \eref{cadas}], so in general the
perturbations cannot be adiabatic.

\item If $c_s^2>0$, the perturbations oscillate.
When we take into account the $\dot{\Delta}$ term, what we have is a
damped harmonic oscillator, and thus the oscillations have decaying
amplitude. As the matter perturbations grow typically as
$\delta_M\sim a$, this ensures that $\delta_D/\delta_M\rightarrow
0$, i.e. that DE will be a smooth component, as usually assumed.
Nevertheless, the larger the scale $l$ or the smaller the speed of sound
$\csd$, the more important DE perturbations are, because then
$k^{-1}\ll \lambda_s$ is not such a good approximation.
\end{itemlist}

Now the question is whether the scales relevant for the matter power
spectrum are really inside the sound horizon or not. The linear
regime of the power spectrum lies in the range
0.01$h$Mpc$^{-1}<k<0.2h$Mpc$^{-1}$ or, equivalently
$(600H_0)^{-1}\lesssim\ell\lesssim(30H_0)^{-1}\label{range}$. On the
other hand, we expect\,\cite{GPS} that at present $\lambda_s\sim
\csd(H_0)^{-1}$. Thus we conclude that (at least for $\csd$ not too
close to 0), the scales relevant for the observations of LSS are
well below the sound horizon, and so the features previously
described apply to them.

Inspection of Eqs. (\ref{ode1}) and (\ref{ode2}) reveals another
important property of the DE perturbations: they diverge if the EOS
of the DE acquires the value $\we = -1$ (known as the ``CC
boundary''), i.e. if the DE changes from quintessence-like (QE)
behavior ($-1/3>\we>-1$) to phantom ($\we<-1$) or vice versa. Note
that, even though $\cad$ diverges at the crossing [cf.
(\ref{cadas})], $(1+\we)c_a^2$ remains finite and, therefore, Eq.
(\ref{ode1}) is well-behaved. Thus, the problem lies exclusively in
the $(1 + \we)$ factor in the denominator of (\ref{ode2}) and only
disappears for vanishing sound of speed $\csd$. One can argue that
the physical source of momentum transfer is not $\theta_D$ but $\rho_D\mathcal{V}_D$, with
$\mathcal{V}_D=\theta_D(1+\we)$\cite{Caldwell05}, and hope to get rid
of the divergence through such a redefinition of variables, but
unfortunately this is not the case. Getting around this difficulty
is not always possible, and in fact there is no way for a single
scalar field (or single fluid) model to cross the CC boundary
\cite{Caldwell05}, and even with two fields some very special
conditions should be arranged. In the absence of a mechanism to
avoid this singularity, we are forced to restrict our parameter
space by removing the points that present such a crossing in the
past.

\section{The $\CC$XCDM model}\label{sec:3}

The properties discussed in the previous section apply in principle
to any model in which the DE is self-conserved. The $\CC$XCDM model,
introduced in Ref.~\refcite{LXCDM12} as a possible explanation to the
cosmological coincidence problem, constitutes a non-trivial example
of these kind of models. In it, the DE is a composite fluid,
constituted by a variable CC and another generic component $X$,
which can exchange energy with $\Lambda$:
\begin{equation}
\rho_D=\rho_{\CC}+\rho_X\,.
\end{equation}
The evolution of $\CC$ can be (as any parameter in QFT) tied to the RG in curved space-time \cite{JHEPCC1}:
\begin{equation}
\frac{{\rm d}\rho_{\CC}}{{\rm d}\ln\mu}=\frac{3\nu}{4\pi}
M_P^2\mu^2\longrightarrow
\rho_\CC=\rho_\CC^0+\frac{3\nu}{8\pi}M_P^2(H^2-H_0^2)\,,\label{lam1}
\end{equation}
where we identified $\mu$ (the energy scale associated to the RG in
Cosmology) with the Hubble function $H$ at any epoch and $\nu$ is a
free, dimensionless, parameter related to the mass ratio (squared)
of the heavy particles contributing to the running versus the Planck
mass\cite{JHEPCC1}. While the ultimate justification for this ansatz
is the application of the RG method and the general considerations
of covariance of the effective action in QFT in curved space-time, a
more profound study is needed, see Refs.~\refcite{SS08} and
\refcite{GPS} (section VI of the latter) for a more detailed
discussion. Interestingly enough, the evolution law (\ref{lam1}) can
be tested from different points of view\cite{RGTypeIa,SS12},
including cosmological perturbations \cite{FabrisDens}(see also
Refs.~\refcite{SB09} and \refcite{Xinmin_perturb05} for related phenomenological
studies).

It should be clear that, in spite of the dynamical nature of $\CC$,
its EOS parameter is $\omega_\CC=-1$, and it is in this sense that
we may call it a ``cosmological constant''. As for the $X$
component, we assume that it has a constant EOS lying in the range
$-1-\delta<\wX<-1/3$ (where $\delta>0$ is small). We need not make
any assumption about the nature of the cosmon, since its evolution
becomes determined  by that of the CC through the energy
conservation equation:
\begin{equation}
\rho'_X+\rho'_\CC+\frac{3}{a}(1+\wX)\rho_X=0\,. \label{lam2}
\end{equation}
The solution of the model in the MD era can be found from Eqs.
(\ref{lam1}), (\ref{lam2}) and the Friedmann equation:
\begin{equation}
H^2=H_0^2[\Omega_M(a)+\OD(a)]\,,\label{fri}
\end{equation}
with $\Omega_M(a)=\Omega_M^0 a^{-3}$. For the normalized DE density,
we find:
\begin{equation}
{\Omega_D}(a)=\frac{{\Omega_{\Lambda}^0}-\nu} {1-\nu}
+\frac{\epsilon\,\Omega^0_{M}\,a^{-3}}{w_X-\epsilon} +
\left[\frac{1-\OLo}{1-\nu}\,
-\frac{w_X\Omega^0_{M}}{w_X-\epsilon}\right]\,a^{-3(1+w_X-\epsilon)}\,,
\label{dewx}
\end{equation}
where we have defined $\epsilon\equiv\nu(1+w_X)$. Assuming as a
prior that $\ODo=\OXo+\OLo\simeq 0.7$, we are left with 3 free
parameters: $\nu$, the parameter that controls the running of
$\CC$; $\omega_X$, the barotropic index of the $X$ component; and
$\OLo$, the current energy density of the CC. Let us note that the
model includes as special cases both the $\CC$CDM ($\nu=0,\,\OXo=0$)
and XCDM ($\nu=0,\,\OLo=0$) models. The effective EOS
parameter of the model,
\begin{equation}\we(a)= -1-\frac{a}{3}
\frac{1}{\Omega_{D}(a)}\frac{d{\Omega_{D}(a)}}{da}\,,
\end{equation}
can present a variety of behaviors\,\cite{LXCDM12} compatible with
$\we(a_0)\simeq -1$ (the subindex 0 standing for the present value),
as suggested by observations\cite{WMAP3}.

\subsection{The coincidence problem}\label{sec:3b}

In order to understand why the $\CC$XCDM model can provide
an explanation for the coincidence problem, it is convenient to
consider the ratio $r$ between the DE and matter energy densities,
which in the standard $\CC$CDM model reads:
\begin{equation}
r\equiv\frac{\OD}{\OM}=\frac{\OLo}{\OMo}a^3\,.
\end{equation}
We see that $r$ tends to zero in the past and grows unboundedly in
the future. Only at the present time we have $r\simeq 1$.
The unavoidable conclusion seems to be that we live in a very
special moment, namely one very close to the time when the expansion
of the Universe started to be accelerated.

In contrast, in the $\CC$XCDM, $r$ reads as follows:
\begin{eqnarray}r&=&\frac{(\OLo-\nu)a^3}{(1-\nu)\,\OMo}+\frac{\epsilon}{\wX-\epsilon}+
\left[\frac{1-\OLo}{\OMo(1-\nu)}-\frac{\wX}{\wX-\epsilon}\right]\,a^{-3\,(\wX-\epsilon)}\,.\label{ratio}
\end{eqnarray}
Such, more complex, structure allows for the existence of a
maximum of this ratio in the future, which implies that $r$ may be
bounded and relatively small (not very different from $1$), say
$r\leq10\,r_0$, for a very prolonged stretch of the history of the
Universe. In this case, the value $r\sim 1$ would no longer be seen
as special.

It is important to note that the ability to solve the coincidence
problem is a very general feature of the model. In order to show
that, let us recall that the solution to the coincidence problem is
linked to the existence of a future stopping of the Universe
expansion\cite{LXCDM12}. Now, from the Friedmann equation
(\ref{fri}), it is clear that it is necessary that the DE density
becomes negative for the expansion to stop. But this condition can
be realized even in the simplest setups of the $\CC$XCDM model. Let
us assume e.g. that $\nu=0$, so there is no exchange of energy
between the CC and the $X$ component:
\begin{equation}
\OD=\OLo+\OXo a^{-3(1+\wX)}\,.
\end{equation}
In this case we have a truly constant $\CC$ and the cosmon behaves effectively as a QE/phantom scalar field; $\OD$
will eventually become negative if any of the following conditions
is fulfilled:
\begin{equation}
\OLo<0 \;\;{\rm and}\;\; -1<\wX<-1/3\qquad {\rm or} \qquad \OXo<0
\;\;{\rm and}\;\; \wX<-1\,.
\end{equation}
Let us also stress that in the $\CC$XCDM the individual components
are not observable, the only thing we can measure is the total
$\OD$. Therefore, there is no problem in having a negative value for
$\OLo$ or $\OXo$, as long as $\ODo=0.7$. Remember also
that $X$ need not be a real fluid, its nature could be effective.

\begin{figure}
\center
\psfig{file=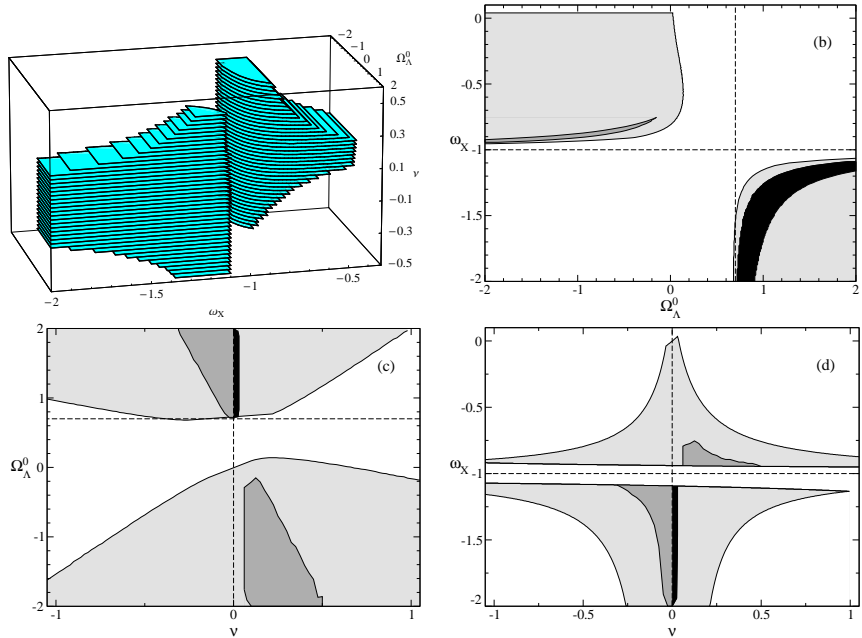,width=0.48\textwidth}\ \ \ \psfig{file=grande1b.eps,width=0.48\textwidth}\\
\psfig{file=grande1c.eps,width=0.48\textwidth}\ \ \
\psfig{file=grande1d.eps,width=0.48\textwidth} \caption{(a) 3D
volume constituted by the points of the $\CC$XCDM parameter space
which provide a solution to the coincidence problem, presenting a
low maximum of the ratio $r$ (\ref{ratio}), $r_{\rm max} \leq 10\,r_0$,
and satisfy the nucleosynthesis bound $r_N\lesssim 0.1$ (see the text); (b), (c)
and (d) Projections of the 3D volume in (a) onto the perpendicular
planes $\nu=0,\,\we=0$ and $\OLo=0$ (all the shaded area). When
we ask for the F-test (\ref{efe}) to be fulfilled and the current
value of the EOS to be close to -1 (\ref{eoslimit}), we are left
with the medium and dark-shaded regions. Finally, by considering
DE perturbations, we are forced to exclude the points for which the
equations are ill-defined, i.e. those for which the EOS of the DE
acquires the value -1 at some point in the past. By doing so we get
the final physical parameter space of the $\CC$XCDM model
(dark-shaded region).}\label{fig:1} \label{figui}
\end{figure}

In \fref{fig:1}a we show that there is a large 3D-volume of the
parameter space for which this solution to the coincidence problem is
possible (the projections of that volume onto three orthogonal planes are shown as the shaded regions in \fref{fig:1}b,\,c,\,d.) All the points in it present a relatively low maximum of the ratio $r$ ($r_{\rm max}\leq10\,r_0)$,
ratio that, in addition, is small enough at the nucleosynthesis
epoch ($r_N\lesssim 0.1$, where in this case $\Omega_M$ is the
density of radiation and $\Omega_D$ is to be computed in the radiation-dominated era), so as to make sure that the predictions of the Big Bang model are not spoiled.

\section{Perturbations in the $\CC$XCDM model}\label{sect:4}

In this section we address the problem of how the parameter space of
a model can be constrained by means of LSS data, using as an example
the $\CC$XCDM model. As we saw in \sref{sec:2b}, at the scales
relevant to the linear part of the matter power spectrum the DE
perturbations are expected to be negligible as compared to the
matter ones. Thus, a reasonable approach is to simply neglect the DE
perturbations from Eq. (\ref{secm}):
\begin{equation}\delta_M''(a)+\frac{3}{2}\left[1-\we (a)\tOD(a)
\right]\frac{\delta_M'(a)}{a}- \frac{3}{2}\,\tOM(a)\
\frac{\delta_M(a)}{a^2}=0\,,\label{linder}\end{equation} and study
the evolution of the perturbations from some initial scale factor
$a=a_i$ in the MD era (where $\delta_M \sim a$) until the present
time, $a_0=1$. As \eref{linder} does not depend on the wave number
$k$, all the scales grow in the same fashion and we can characterize
models by means of the ``growth factor'':
\begin{equation}
D(a)=\frac{\delta_M(a)}{\delta_{M}^{\mbox{\tiny CDM}}(a_0)}\,,
\end{equation}
whose present value, $D(a_0)$, compares the growth of the
perturbations in the model considered to the growth in a pure cold
dark matter (CDM) model. The parameter that measures the agreement
between the observed galaxy distribution power spectrum,
$P_{GG}(k)$, and the matter power spectrum of a model, $P(k)\equiv
|\delta_M(k)|^2$, is the linear bias, which at the present time is
defined as $b^2(a_0)=P_{GG}/P$. Most remarkably, the LSS data point
to the value $b_{\CC}^2(a_0)=1$, to within a 10\% accuracy, for the
$\CC$CDM model\cite{Cole05}. This suggests that the comparison to
the $\CC$CDM can be a valid and more economical criterion for
studying the viability of a model. In particular, we may require
that any DE model should pass the following
``F-test''\cite{Ftest}\,:
\begin{equation}
|F|\equiv \left|1-\frac{b^2(a_0)}{b^2_\CC(a_0)}\right|
=\left|1-\frac{P_\CC(a_0)}{P(a_0)}\right|=\left|1-\frac{D^2_\CC(a_0)}{D^2(a_0)}\right|\leq
0.1\,.\label{efe}
\end{equation}
This was done for the $\CC$XCDM model (and also for the running CC model\,\cite{JHEPCC1})
in Ref.~\refcite{GOPS}, where, in addition, we imposed
that the current value of the EOS parameter of the DE should be
close to -1:
\begin{equation}
|\we(a_0)+1|\leq 0.3\,, \label{eoslimit}
\end{equation}
as suggested by recent observational
limits\,\cite{WMAP3}$^,$\footnote{Let us remark, though, that such limits on
the EOS parameter are usually derived under the assumption of a
constant $\we$ and, therefore, do not strictly apply to our model.}.
As seen in \fref{fig:1}b,\,c,\,d, there is still a sizable region of
the parameter space (medium and dark-shaded regions) where the
$\CC$XCDM satisfies these two new conditions and the nucleosynthesis
bound, while still providing a solution to the coincidence problem.

Neglecting DE perturbations provides us therefore with a simple and
effective method to constrain the parameter space of a model.
Although we expect it to be a reasonable approximation, we cannot be
completely sure unless we perform a full analysis in which the DE
fluctuations are also included. Such an analysis\,\cite{GPS} implies
an immediate and very important consequence. As discussed in
\sref{sec:2b}, if the effective EOS of the model crosses the CC
boundary ($\we=-1$) at some point in the past, the perturbation
equations will diverge. In the absence of a mechanism to get around
this singularity (and indeed we cannot have it without a microscopic
definition of the $X$ component, i.e. one that goes beyond a mere
conservation law), we are forced to restrict our parameter space by
removing the points that present such a crossing. This new
constraint knocks off many of the points allowed by the previous
simple analysis; in fact, we are left with the dark-shaded region in
\fref{fig:1}b,\,c,\,d, and so we end up with a rather definite
prediction for the values of the parameters of the $\CC$XCDM model.
It is worth noticing that only small (and positive) values of $\nu$
are allowed, $\nu\sim10^{-2}$ at most, which is in very good
agreement with theoretical expectations\cite{GPS}. Another
interesting consequence of the new constraint is that the effective
EOS of the DE can be QE-like only\cite{GPS}, i.e. $-1<\we<-1/3$.

We want to compare the matter power spectrum predicted by the
$\CC$XCDM model, $P_{\CC X}(k)$, with the $P_{GG}(k)$\footnote{And
also with the $\CC$CDM spectrum, $P_{\CC}(k)$, which provides a good
fit to $P_{GG}(k)$} measured by the 2dFGRS
collaboration\cite{Cole05}. The former can be found by evolving
the perturbation equations (\ref{ode})-(\ref{ode2}) from $a=a_i$ to
$a_0=1$, where in this case $a_i\ll 1$ is the scale factor at some
time well after recombination. In order to set the initial
conditions, we took into account that the DE does not begin to play
an important role until very recently, so that the values of the
metric and matter perturbations at $a=a_i$ should be the same for
our model and for the $\CC$CDM model -- the power spectrum
$P_{\CC}(k)$ of the latter being available from standard analytical
fits in the literature, see Ref.~\refcite{GPS} and references therein.
As for the DE perturbations, we assumed that they vanish at $a=a_i$.
This is reasonable because, as noticed before, the DE perturbations
are expected to be negligible at the scales relevant to the linear
part of the matter power spectrum.

\begin{figure}
\center \psfig{file=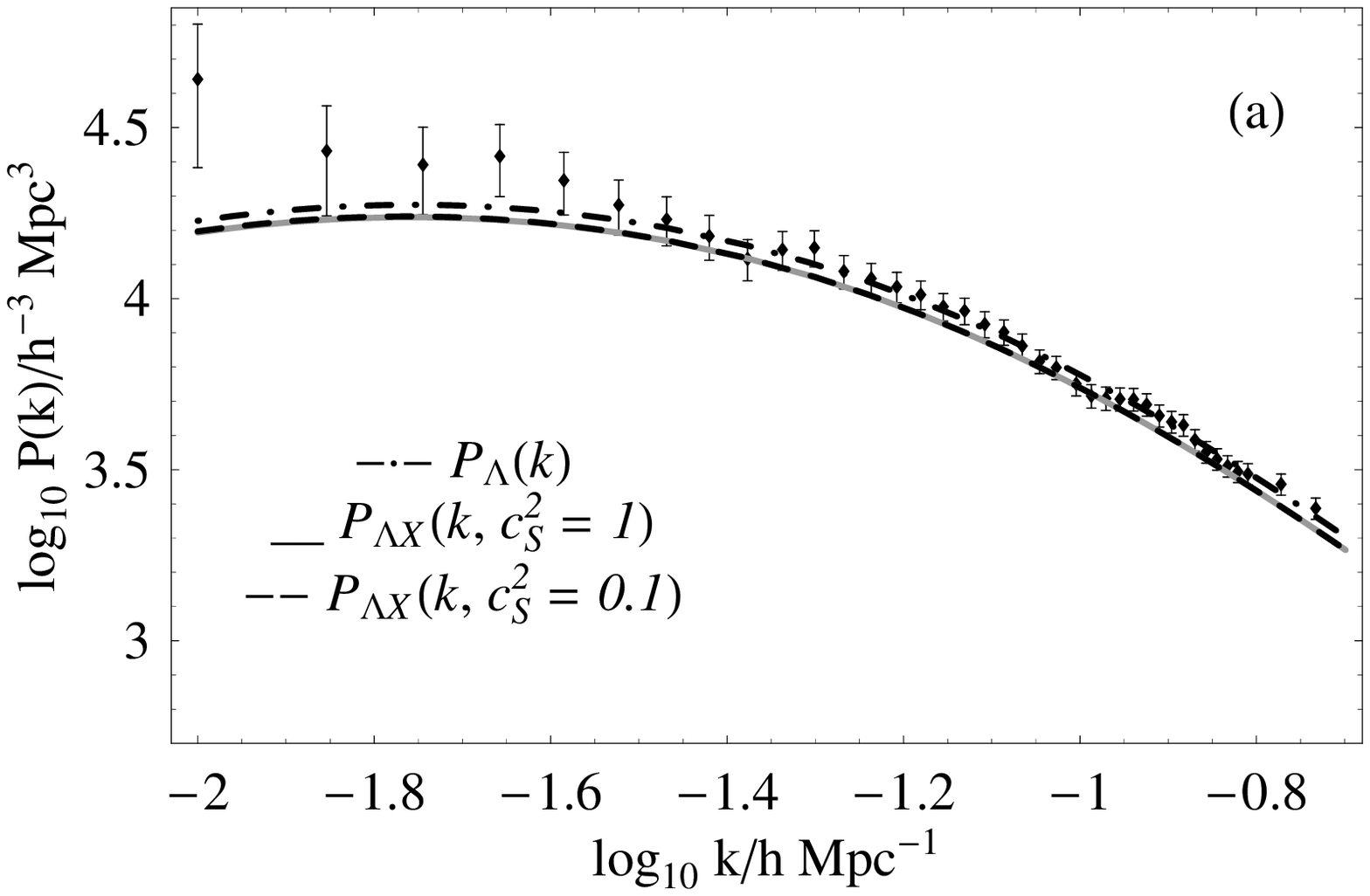,width=0.495\textwidth}
\psfig{file=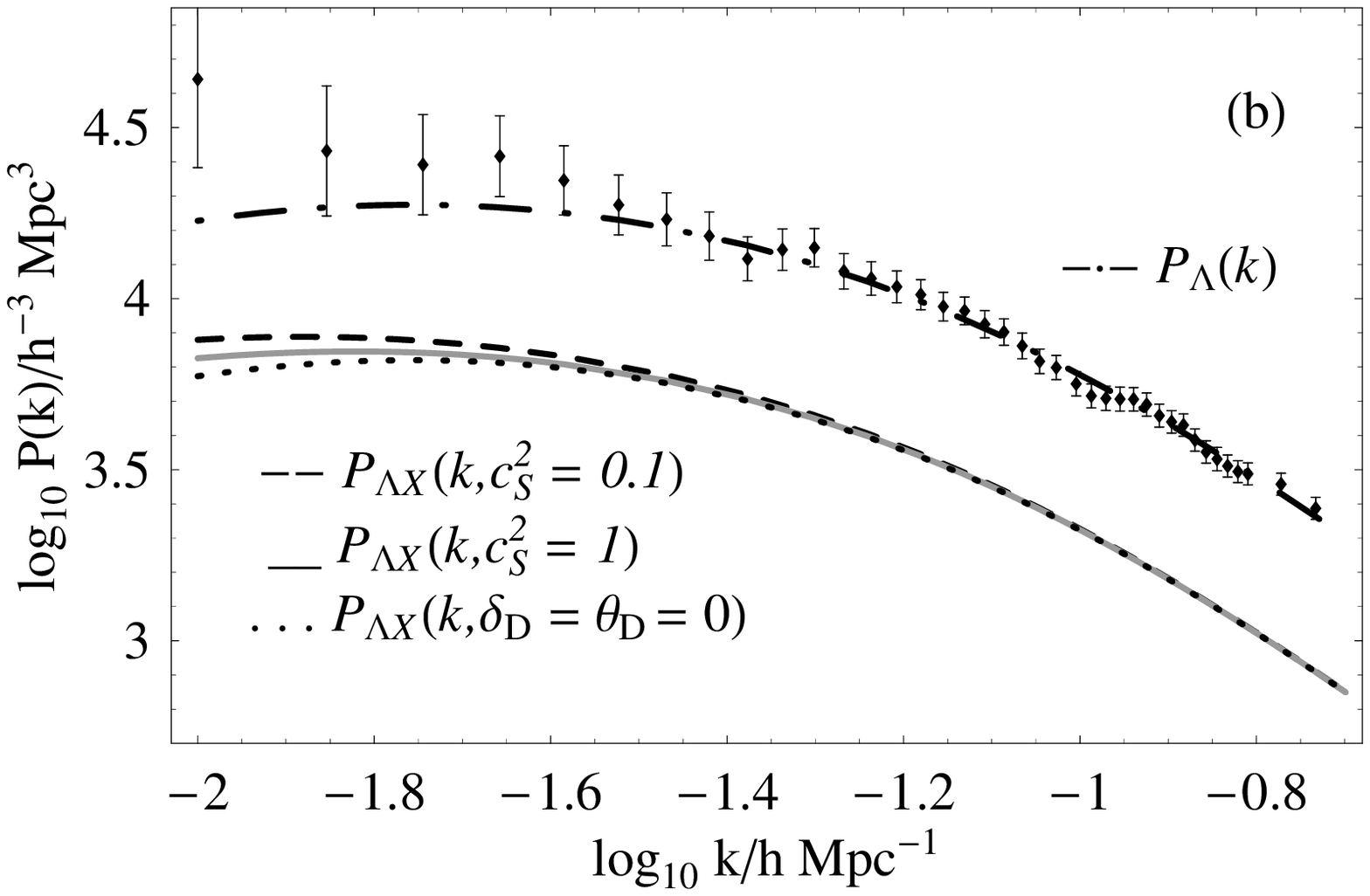,width=0.495\textwidth} \caption{The 2dFGRS
observed galaxy power spectrum\cite{Cole05}, $P_{GG}(k)$ (points),
and the $\CC$CDM power spectrum, $P_\CC(k)$ (dot-dashed line) versus
the spectrum predicted by the $\CC$XCDM, $P_{\CC X}(k)$, for DE
sound speeds $\csd=0.1$ (dashed line) and $\csd=1$ (solid/gray
line): (a) for a set of parameters allowed by the analysis of
Ref.\,\refcite{GOPS} (in the dark-shaded region of
\fref{fig:1}b,\,c,\,d), $\Omega_{\Lambda}^0 =0.8$, $\nu =
\nu_0\equiv 2.6\times 10^{-2}$ and $w_X=-1.6$; (b) for a set of
parameters satisfying all the conditions in that analysis but the
F-test, $\Omega_{\Lambda}^0 = +0.35$, $\nu =-0.2$ and $w_X=-0.6$. In
this case it is also shown the power spectrum obtained by neglecting
DE perturbations (dotted line), which presents the same shape as
$P_\CC(k)$.}\label{fig:2}
\end{figure}

The $\CC$XCDM power spectrum was calculated for two different
fiducial sound speeds, $c_{s}^2=1$ and $c_{s}^2=0.1$ and several
combinations of the parameters $\nu$, $\wX$ and $\OLo$. For values
of the parameters \textit{not} fulfilling the F-test (even though
satisfying all the other conditions stated in \fref{fig:1}) we
obtain huge discrepancies, as expected. The discrepancy appears as
an approximate global suppression gap (in the entire $k$ range) of
the amount of growth with respect to the $\CC$CDM model (cf.
\fref{fig:2}b). This suppression is typical of the QE-like
behavior\,\cite{GPS} and occurs even if the DE perturbations are
neglected (dotted line), in which case $P_{\CC X}(k)$ presents
exactly the same shape as $P_{\CC}(k)$ (because then the
$k$-dependence disappears from the equations, which reduce just to
\eref{linder}). The effect of considering DE perturbations is only
visible at large scales (small $k$), where they tend to compensate
the aforementioned suppression. The smaller the speed of sound or
the larger the scale, the more important is the effect of DE
perturbations, as expected from the general considerations of
\sref{sec:2b}.

\begin{figure}
\psfig{file=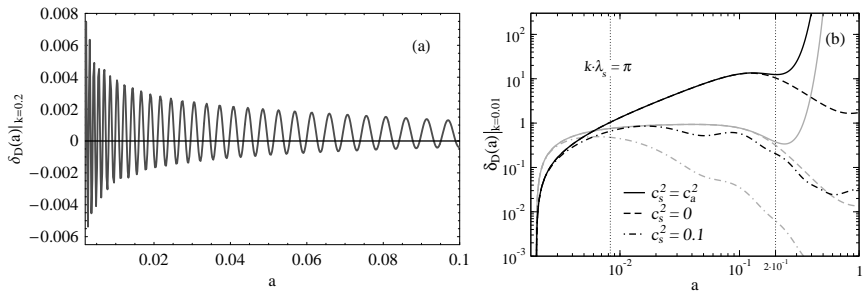,width=0.545\textwidth}
\psfig{file=grande3b.eps,width=0.445\textwidth} \caption{(a) The
$\CC$XCDM growth of DE perturbations for a small scale $k=0.2$ (in
units of $h\,$Mpc$^{-1}$) and the same set of parameters assumed in
\fref{fig:2}a, and for DE sound speed $c_{s}^2=0.1$; (b) Evolution
of the DE perturbations $\delta_D$ (black lines) for the same set of
parameters as in (a), at the large scale $k=0.01$ and for three
different speeds of sound: $\csd=\cad<0$ (solid line), $\csd=0$
(dashed line) and $\csd=0.1$ (dot-dashed line). The evolution of the
ratio $\delta_D/\delta_M$ is also shown (gray lines).} \label{fig:3}
\end{figure}

In contrast, in \fref{fig:2}a we see that for values allowed by the
F-test (and satisfying all the other constraints as well, i.e. lying
in the dark-shaded region in \fref{fig:1}b,\,c,\,d), $P_{\CC X}(k)$
is very similar to $P_{\CC}(k)$, with numerical results in very good
agreement with those obtained through the F-method\cite{GOPS,GPS}.
In particular, their shape is identical, indicating that DE
perturbations do not play a role here. Indeed, in \fref{fig:3}a we
see that $\delta_D$ oscillates with decreasing amplitude, as
predicted in \sref{sec:2b}. For positive sound speed, the
perturbations get stabilized (and therefore the ratio
$\delta_D/\delta_M$ becomes negligible) once the sound horizon
(\ref{po}) is crossed, i.e. when $k\lambda_s=\pi$, as seen in
\fref{fig:3}b. Similarly, in the adiabatic case, the perturbations
begin their exponential growth once $\cad$ (which is negligible in
the far past in the $\CC$XCDM model\,\cite{LXCDM12}) eventually
takes a sizable negative value. The runaway behavior is triggered
by the term proportional to $k^2\,\cad<0$ in (\ref{ode2}), or
equivalently in (\ref{secd}).

\section{Conclusions}

We have analyzed the behavior of the DE perturbations in models with
self-conserved DE. We have exemplified them by means of the
$\CC$XCDM model\cite{LXCDM12}, which is a non-trivial model of the
cosmic evolution with a number of appealing properties. Unlike other
proposed solutions to the coincidence problem (an incomplete list
includes tracking scalar fields, interactive QE models, K-essence,
Chaplygin gas, etc - see e.g. Ref.~\refcite{GPS} and references
therein), the $\CC$XCDM model accounts for the energy of vacuum
through a (possibly running) $\CC$, giving allowance for other
dynamical contributions, $X$, of general nature. The comparison of
the $\CC$XCDM power spectrum to the LSS data, first by means of the
F-test and then through a full analysis of the DE perturbations,
resulted in a strong additional constraint on the parameter space of
the model, hence increasing its predictive power and pinpointing a
region where the $\CC$XCDM model provides a realistic solution to
the coincidence problem, i.e. fully compatible with present
observations.

\section*{Acknowledgments}
Authors have been supported in part by MEC and FEDER under project
FPA2007-66665 and also by DURSI Generalitat de Catalunya under
project 2005SGR00564. We acknowledge the support from the Spanish
Consolider-Ingenio 2010 program CPAN CSD2007-00042. JG is also
supported by MEC under BES-2005-7803 and is grateful to the Institut f\"ur
Theoretische Physik, Universit\"at Heidelberg, for the hospitality and support.

\end{document}